\documentstyle[12pt]{article}
\begin{document}
\twocolumn
\twocolumn[\baselineskip=0.425cm
~ \par
\begin{center}
{\large\bf Tunneling conductance of normal 
metal/insulator/Sr$_{2}$RuO$_{4}$ junction}\par
\vspace{4.5mm}
{\small Masashi Yamashiro and Yukio Tanaka } \par
\vspace{3mm}
{\small Graduate school of Science and Technology, and 
Department of Physics, Niigata University, 
Ikarashi Niigata 950-21, Japan} \par
\vspace{3mm} 
{\small Satoshi Kashiwaya } \par
\vspace{3mm}
{\small Electrotechnical Laboratory, Tsukuba, Ibaraki 
305, Japan} \par
\end{center}
\vspace{3mm} 
\noindent
{\small \ \ \
A theory of tunneling conductance spectra for normal 
metal/insulator/Sr$_{2}$RuO$_{4}$ junction is studied theoretically.
We assume several types of pair potentials with 
triplet symmetries that are promising candidates for Sr$_{2}$RuO$_{4}$.
The calculated conductance spectra
show either zero bias peaks or gap like structures 
depending on the orientations of the junctions. 
The existence of a residual components in the gap 
reflects a non-unitary property of the superconducting states. 
Based on the calculation of a temperature dependence,
we will verify that the measurement temperature below 100mK is required
to determine the pairing symmetry experimentally.  } \par
\vspace{7.0mm}]
\small
\baselineskip=0.425cm
\noindent
{\bf 1. Introduction} \par
Recently, Maeno $et$. $al$.\cite{maeno} 
discovered a superconducting state in Sr$_{2}$RuO$_{4}$. 
Since this material has layered perovskite structure, 
strong anisotropic properties in electronic states are expected 
for both normal and superconducting state\cite{yoshida}. 
Several experiments indicate that
rather large residual density of states remain even at low 
temperatures
\cite{maeno,ishida}. 
Moreover, the existence of the ferromagnetic spin
fluctuations is 
supported by several evidences
\cite{rice,mazin}. 
Based on these facts, 
two-dimensional nonunitary 
triplet superconducting states 
which belong 
to two-dimensional odd-parity E$_{u}$ 
representation under the tetragonal 
symmetry have been suggested
theoretically
\cite{sigrist,machida}. 
However, a definitive evidence for the triplet symmetry states 
has not been presented so far. 
Therefore, more precise and detailed experiments are expected
to get a clear conclusion.\par
Tunneling spectroscopy is one of the candidates for such experiments
because of its
high energy resolution.
Moreover a phase sensitive capability of tunneling spectroscopy
has been revealed \cite{tanaka,kashiwaya}.
However, no theory exists for tunneling spectroscopy
for triplet superconductors even now.
In this paper, we study the tunneling conductance spectra
for 
normal metal / insulator / Sr$_{2}$RuO$_{4}$ (N/I/S)
junction 
in the finite temperature region by extending the previous
theories
\cite{tanaka,kashiwaya,yamashiro}. 
The transition temperature T$_{\mbox{c}}$ of 1.04K is assumed
in the following. \par
\vspace{0.43cm}
\noindent
{\bf 2. Model} \par
For the calculation of the tunneling conductance, 
we assume an N/I/S junction model in the clean limit 
with a semi-infinite structure. 
A nearly two-dimensional Fermi surface is assumed. 
We calculate the conductance for two kinds of orientations with completely 
flat interfaces; one is perpendicular to 
the $x$-axis and the other is perpendicular to 
the $z$-axis. 
The Fermi wave number and the effective mass are assumed
to be equal both in the normal metal and in the
superconductor. 
\par
Hereafter,
following the discussion by Sigrist\cite{sigrist} and
Machida
\cite{machida}, 
we will choose two kinds of nonunitary pair potentials with 
tetragonal symmetry. 
The pair potential forms are
$\Delta_{\uparrow\uparrow}=\Delta_{0}$
$\times\sin\theta(\sin\phi+\cos\phi)$ 
and $\Delta_{\uparrow\uparrow}=\Delta_{0}e^{i\phi}\sin\theta$, 
that are referred to as 
E$_{u}$(1) and E$_{u}$(2), respectively.
Here 
$\Delta_{\uparrow\uparrow}$ is $(\uparrow$, $\uparrow)$
component of 
the pair potential matrix in the spin space, and 
$\Delta_{0}$ is the amplitude of the pair potential in
the bulk state.
The details of the calculation for the conductance formula 
and results at zero temperature are already discussed in
Ref.\cite{yamashiro}.
Here, we extend previous results to finite
temperature cases. 
The conductance formula for $z$-$y$ plane interface 
is given by
\[
\begin{array}{l}
\displaystyle
\sigma_{S}(eV)\propto\frac1T
\int_{-\infty}^{\infty}\!
\int_{\pi/2-\delta}^{\pi/2}\!\int_{-\pi/2}^{\pi/2}
(\sigma_{S,\uparrow}+\sigma_{S,\downarrow})\sigma_{N} \\
\noalign{\vskip 2ex}
\displaystyle
\hspace{22pt}\times\mbox{sech}^{2}
\left( \frac{E-eV}{2k_{B}T} \right)
\sin^{2}\theta\cos\phi d\theta d\phi dE \\
\noalign{\vskip 2ex}
\displaystyle
\sigma_{N}(eV)\propto\frac2T
\int_{-\infty}^{\infty}\!\int_{\pi/2-\delta}^{\pi/2}
\!\int_{-\pi/2}^{\pi/2}\sigma_{N}
\mbox{sech}^{2}\left( \frac{E-eV}{2k_{B}T} \right) \\
\noalign{\vskip 2ex}
\displaystyle
\hspace{22pt}\times\sin^{2}\theta \cos\phi d\theta d\phi dE 

\end{array}
\]
\begin{equation}
\displaystyle
\sigma(eV)=\frac{\sigma_{S}(eV)}{\sigma_{N}(eV)}
\label{eqn:e1}
\end{equation}
The quantity $\sigma_{S,\uparrow(\downarrow)}$ is obtained 
in a similar way  at zero temperature using the coefficients
of 
the Andreev reflection and normal reflection, 
where $\sigma_{N}$ is the tunneling conductance in the
normal state. 
The quantity $\delta$ expresses the degree of 
the two-dimensionality of the Fermi surface which is chosen as $0.05\pi$.
The conductance formula for the $x$-$y$ plane interface is 
obtained in a similar way. 
\par
\vspace{0.43cm}
\noindent
{\bf 3. Results} \par
The calculated conductance spectra
$\sigma(eV)$ are 
plotted for various temperatures and barrier heights. 
For all cases, due to the nonunitary 
property of the pair potential, 
$\sigma(eV)$ is definitely larger than 0.5. 
As seen from 
[Fig.(\ref{fig:f1})] to [Fig.(\ref{fig:f4})], 
with the increase of temperature, 
$\sigma(eV)$ becomes nearly constant. 
The temperature dependence is significant  
in the case of $Z=5$ both for $E_{u}$(1) 
and $E_{u}$(2) states. 
For E$_{u}$(1) state, $\sigma(0)$ is enhanced at low
temperatures 
in the $z$-$y$ plane junctions as shown in Fig.(\ref{fig:f2}). 
In this direction, 
$\sigma(0)$ of E$_{u}$(2) state converges to a nonzero value
at zero temperature. 
This difference is due to the fact whether zero energy
states 
at the interface are formed on the finite area of the Fermi
surface 
or the line of the Fermi surface. 
In the large $Z$ limit, 
zero energy states (ZES)
are expected for $-\pi/4 \leq \phi \leq \pi/4$ in the case
of 
E$_{u}$(1) state. 
While ZES are expected for $\phi=0$ in the case of
E$_{u}$(2) state. 

In the case of $x$-$y$ plane interface junction, 
gap-like spectra are obtained. 
The difference of the conductance spectra between E$_{u}$(1)
state and 
E$_{u}$(2) state is clearly seen below $T \leq 0.1T_{C}$ and
Z=5. 
\par
In conclusion, to determine the symmetry of pair potential
in Sr$_{2}$RuO$_{4}$,
tunneling spectroscopy measurements
will give us definitive information
if the measurements will be
performed at a temperature lower than 100 mK 
and by using highly controlled junctions.
\par

\vspace{0.43cm}
\noindent
{\bf Acknowledgments} \par
One of the authors (Y.T.) is supported by a Grant-in-Aid for
Scientific 
Research in Priority Areas, ``Anomalous metallic states near

the Mott transition'', and ``Nissan Science Foundation''. 
The computation in this work has been done using the 
facilities of the Supercomputer Center, Institute for Solid 
State Physics,  University of Tokyo. \par
\vspace{0.43cm}
\par

\begin{figure}
\caption{$\sigma(eV)$ is plotted as a function of $eV /
\Delta_{0}$,
with a:T=0.1T$_{c}$, b:T=0.3T$_{c}$, and c:T=0.8T$_{c}$.}
\label{fig:f1}
\end{figure}
\begin{figure}
\caption{$\sigma(eV)$ is plotted as a function of $eV /
\Delta_{0}$, 
with a:T=0.1T$_{c}$, b:T=0.3T$_{c}$, and c:T=0.8T$_{c}$.}
\label{fig:f2}
\end{figure}
\begin{figure}
\caption{$\sigma(eV)$ is plotted as a function of $eV /
\Delta_{0}$, 
with a:T=0.1T$_{c}$, b:T=0.3T$_{c}$, and c:T=0.8T$_{c}$.}
\label{fig:f3}
\end{figure}
\begin{figure}
\caption{$\sigma(eV)$ is plotted as a function of $eV /
\Delta_{0}$,
with a:T=0.1T$_{c}$, b:T=0.3T$_{c}$, and c:T=0.8T$_{c}$.}
\label{fig:f4}
\end{figure}

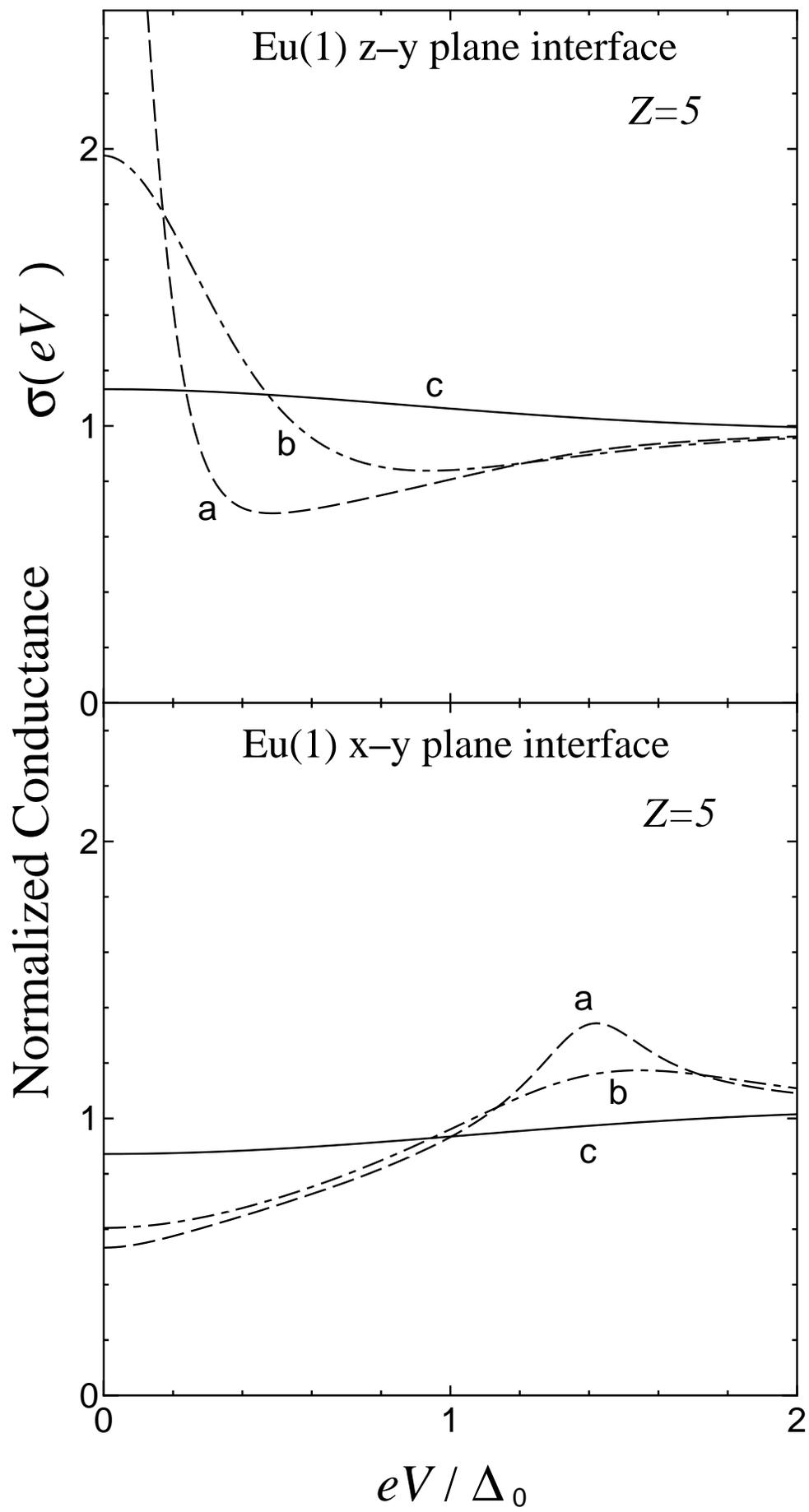
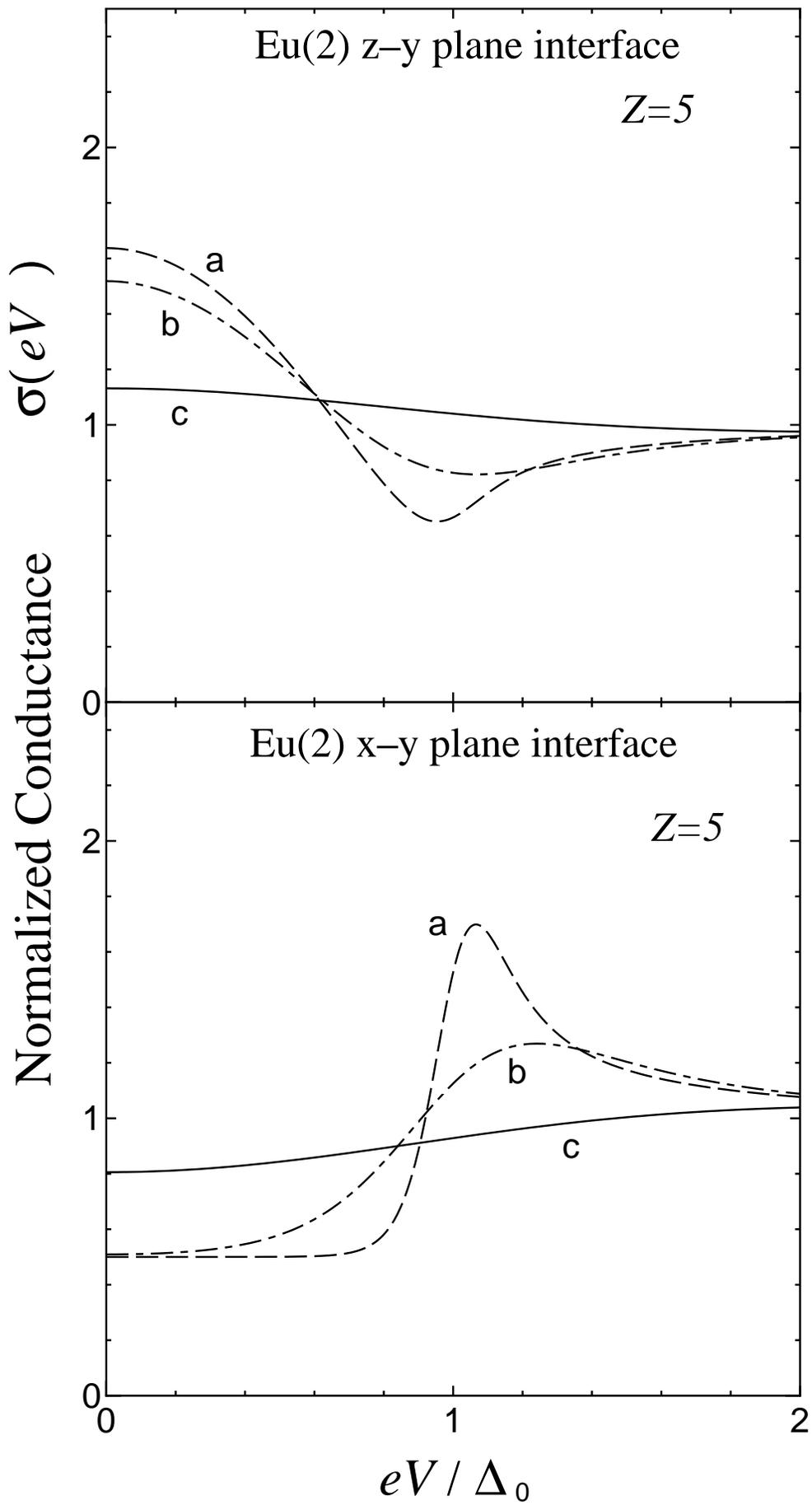
\end{document}